# In-plane angular dependence of the spin-wave nonreciprocity of an ultrathin film with Dzyaloshinskii-Moriya interaction

Vanessa Li Zhang, Kai Di, Hock Siah Lim, Ser Choon Ng, Meng Hau Kuok[a]

*Department of Physics, National University of Singapore, Singapore 117551*

Jiawei Yu, Jungbum Yoon, Xuepeng Qiu, and Hyunsoo Yang[b]

*Department of Electrical and Computer Engineering, National University of Singapore, Singapore 117576*

**Abstract**

**The nonreciprocal propagation of spin waves in an ultrathin Pt/Co/Ni film has been measured by Brillouin light scattering. The frequency nonreciprocity, due to the interfacial Dzyaloshinskii-Moriya interaction (DMI), has a sinusoidal dependence on the in-plane angle between the magnon wavevector and the applied magnetic field. The results, which are in good agreement with analytical predictions reported earlier, yield a value of the DMI constant which is the same as that obtained previously from a study of the magnon dispersion relations. We have demonstrated that our magnon-dynamics based method can experimentally ascertain the DMI constant of multilayer thin films.**

[a] phykmh@nus.edu.sg.

[b] eleyang@nus.edu.sg.

Because of its interesting fundamental physics and enormous potential applications, the Dzyaloshinskii-Moriya interaction (DMI) has attracted extensive scientific attention lately. An antisymmetric exchange interaction, the DMI is a crucial mechanism in novel magnetic phenomena such as molecular magnetism [1], the magnon Hall effect [2], and the magnetically induced electric polarization in multiferroics [3]. Most importantly, the DMI is responsible for the formation of the magnetic skyrmion, a chiral spin texture which was discovered in 2009 [4-6]. This remarkable magnetic entity is a potential information carrier in next-generation low-energy, ultrahigh-density magnetic storage devices, owing to its special attributes like minute size, propagation under ultralow current densities [7,8] and rewritability by spin-polarized currents [9].

First proposed to exist in non-centrosymmetric bulk materials by Moriya [10], the DMI has also been found at the interfaces between magnetic films and high spin-orbit metals [11-13]. Interfacial DMIs were also observed in multilayer thin films comprising a magnetic layer sandwiched between a strong spin-orbit metal and an oxide layer [14-17]. These interactions can exist in such multi-component structures as, due to the different under- and over-layers, they do not possess inversion symmetry.

Because of their significance (e.g. skyrmions in magnetic materials lacking inversion symmetry are stabilized by chiral DMIs [18]), it is of importance to experimentally determine the strength of these interactions, which is represented by the DMI constant. The spin-polarized electron energy loss spectroscopic [13] and inelastic neutron scattering [19] techniques are available for directly probing DMIs, but they are not suitable for multilayers with buried metal/ferromagnet thin films [20]. Brillouin light scattering (BLS) [21], in contrast, is particularly suited for studying magnons, with frequencies in the gigahertz range, in such film stacks. Di *et al*. [16,17] have very recently demonstrated that BLS offers a sensitive and convenient way for the direct observation of interfacial DMIs in multilayer films, such as MgO/Pt/Co/Ni/MgO/SiO$_2$ and MgO/Pt/CoFeB/MgO/SiO$_2$. They found that the interactions were manifested as the asymmetry of the measured dispersion relations of counter-propagating spin waves (SWs) in the films. An analytical theory for the dispersion relations of SWs in ferromagnetic films with DMIs has recently been formulated by Cortés-Ortuño and Landeros [22]. Here, we report on an experimental study of the dependence of the spin-wave nonreciprocity of a Pt/Co/Ni film on the in-plane angle between the spin-wave wavevector and the applied magnetic field. The Brillouin data obtained were used to verify their theory on the angular dependence of the spin-wave nonreciprocity.

The sample investigated was an unannealed film stack, substrate/MgO(2)/Pt(4)/Co(1.6)/Ni(1.6)/MgO(2)/SiO2(3), where the figures in parentheses are the respective thicknesses in nm. It was deposited on a thermally oxidized silicon wafer by both DC and RF magnetron sputtering at room temperature. Details of the fabrication procedure can be found in Ref. 16. Vibrating sample magnetometer measurements of the magnetic hysteresis loops of the Pt/Co/Ni sample gave the in-plane saturation field and the saturation magnetization $M_S$ as 50 mT/$\mu_0$ (where $\mu_0$ is the permeability of free space) and 1160 kA/m, respectively.

The Brillouin experiments were carried out in the 180° back-scattering geometry using the 514.5 nm radiation of an argon-ion laser and a six-pass tandem Fabry-Perot interferometer. The magnon wavevector ***k*** lies along the intersection of the scattering plane (shaded orange) and the film plane (*x-z* plane), as illustrated in Fig. 1. All measurements were made with the incident angle of the laser beam set at $\theta = 30°$, and thus with the wavevector fixed at $k$ (= $4\pi\sin\theta /514.5$) = 12.2 μm$^{-1}$. The sample was subjected to an in-plane field of $H_0$ = 64 mT/$\mu_0$ by symmetrically positioning it between a pair of oppositely-poled disc permanent magnets. The sample and the magnet pair were mounted on a miniature two-axis translation stage attached to a graduated rotation stage which permitted variation in the in-plane angle $\phi$ between the field $H_0$ (fixed in the $z$ direction) and the magnon wavevector $k$ (see Fig. 1). This stage was, in turn, attached to a vertically mounted rotation stage, with their axes orthogonal, allowing $\theta$ to be set at 30°. This composite sample holder ensured that the laser light irradiated the same spot on the film for any setting of $\phi$. Note that for the isotropic ferromagnetic film studied, and with $H_0$ > 50 mT/$\mu_0$ (the in-plane saturation field), the saturation magnetization $M_S$ is parallel to $H_0$, and thus $\phi$ is also the angle between $M_S$ and $k$.

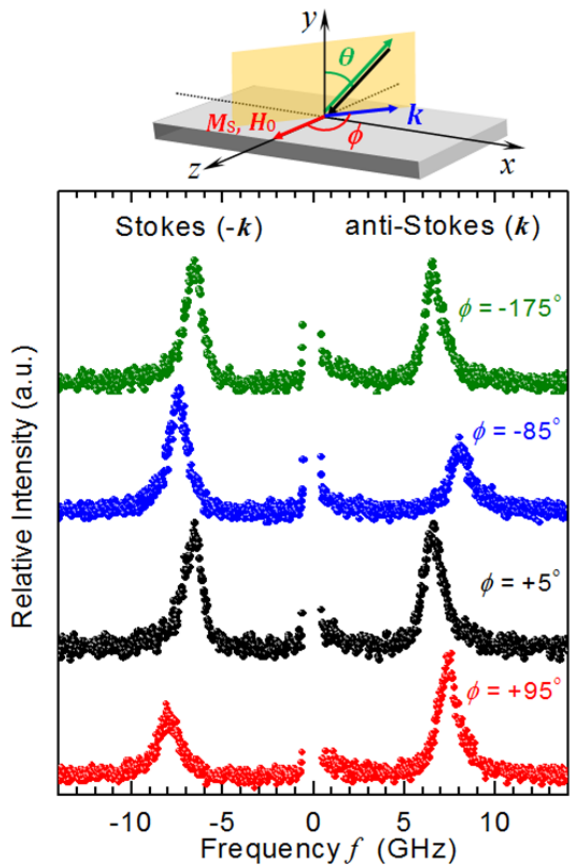

**FIG. 1.** Top: Schematic of the 180°-backscattering geometry used, showing the in-plane angle $\phi$ between the magnon wavevector $\mathbf{k}$ and applied magnetic field $\mathbf{H}_0$, with both vectors lying in the x-z plane of the Pt/Co/Ni film. For the illustrated case, $-90° < \phi < -180°$. The incident and scattered light beams, denoted by respective black and green arrows, lie in the scattering plane (shaded orange) and are at incident angle $\theta = 30°$. The saturation magnetization $\mathbf{M}_S$ and $\mathbf{H}_0$ are in the z direction. Bottom: Brillouin spectra of the Pt/Co/Ni film measured under $H_0 = 64$ mT/$\mu_0$, and for various in-plane angles $\phi$.

Brillouin spectra were recorded for values of $\phi$ in the $-180° \leq \phi \leq 180°$ range, and several typical spectra are presented in Fig. 1. The thinness of the metallic film permitted the simultaneous observation of counter-propagating surface SWs localized at the top and bottom interfaces. Each corresponding pair of SWs, propagating in the $-k$ and $+k$ directions, would appear as the respective Stokes and anti-Stokes peaks in a Brillouin spectrum. Except for the spectra measured in the vicinities of $\phi = -180°$, $0°$ and $180°$, the frequencies of the Stokes and anti-Stokes peaks are different. This observed asymmetry, which is maximal in the vicinities of $\phi = -90°$ and $90°$, is indicative of the most pronounced nonreciprocity in the SW propagation. Figure 2 presents the dependence of the frequencies $f$ of the Stokes and anti-Stokes peaks on the in-plane angle $\phi$. There is a clear periodic trend in the variation. In the $-180° < \phi < 0°$ range, the Stoke frequencies are lower than those of the anti-Stokes, while the reverse is true for $0° < \phi < 180°$, indicating an obvious angular variation of the SW nonreciprocity.

Using the micromagnetic theory, Cortés-Ortuño and Landeros have established a general dispersion relation for thin films possessing DMI of various structural symmetries [22]. For our isotropic film, its saturation magnetization is parallel to the applied field $H_0$ which is fixed along the $z$ axis, and the wavevector $k$ lies in the $x$-$z$ film plane. The DMI energy density is given by $\frac{D}{M_S^2}\left(M_x\frac{\partial M_y}{\partial x} - M_y\frac{\partial M_x}{\partial x} + M_z\frac{\partial M_y}{\partial z} - M_y\frac{\partial M_z}{\partial z}\right)$ [18,23,24], where $D$ is the DMI constant, and $M_i$ the $i$-component of the magnetization. Following Ref. 22, the dispersion relation can be written as

$$2\pi f(k) = \frac{2\gamma}{M_S}Dk\sin\phi + \mu_0\gamma\sqrt{\left[H_0 + Jk^2 + F(kd)M_S\sin^2\phi\right]\left[H_0 + M_S - H_U + Jk^2 - F(kd)M_S\right]},$$

(1)

where $\gamma$ is the gyromagnetic ratio, $d$ the film thickness, $F(x) = 1 - (1 - e^{-|x|})/|x|$, the uniaxial anisotropic field $H_U = 2K/(\mu_0 M_S)$, $K$ being the anisotropy constant, and $J = 2A/(\mu_0 M_S)$, $A$ being the exchange stiffness constant. It is noted that only the first term on the right hand side, which is linear in $k\sin\phi$, depends on the DMI. It is responsible for the nonreciprocal propagation along $k$ and $-k$, and is maximal for $\phi = \pm 90°$, at which value the saturation magnetization lies in the film plane and is normal to $k$. The second term on the right hand side, which is symmetric with respect to wavevector inversion, represents the magnon dispersion for magnetic films in the absence of DMI.

The respective dependences of the experimental frequencies of counter-propagating spin waves, obtained from corresponding Stokes and anti-Stokes peaks, on $\phi$ were fitted to Eq. (1), yielding an interfacial DMI constant value $D = 0.44$ mJ/m$^2$. The good agreement between experiment and theory can be seen in Fig. 2. It should be pointed out that the values of the magnetic parameters used in the fitting were set to those reported by Di *et al.* for Pt/Co/Ni [16], namely $M_S$ = 1160 kA/m, $H_U$ = 670 kA/m, $\gamma$ = 194 GHz/T, and $J$ = 2.33 × $10^{-11}$ A·m. The calculated angular dependence corresponding to the absence of the DMI, shown in Fig. 2, is symmetric about $\phi$ = 0°, meaning that for such a situation, the SW propagation is reciprocal in nature.

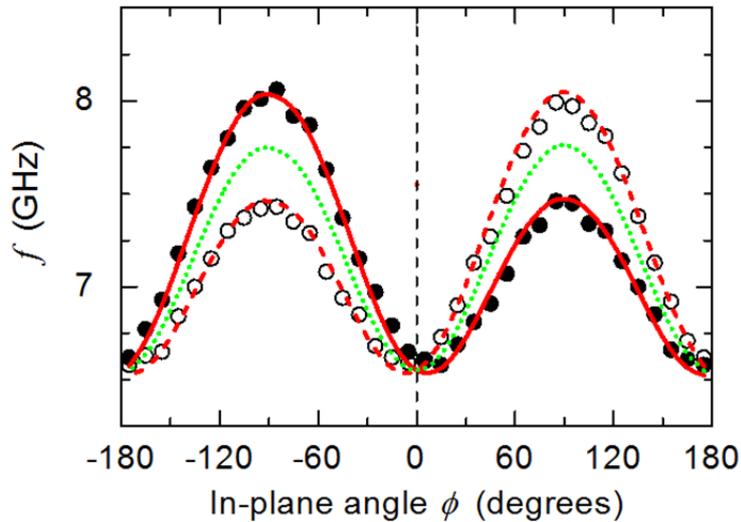

**FIG. 2.** Dependences of the respective frequencies of counter-propagating spin waves of the Pt/Co/Ni film on the in-plane angle $\phi$ between the wavevector $k$ and the applied magnetic field $H_0$. Open and closed circles denote the respective measured frequencies of the Stokes ($-k$) and anti-Stokes ($k$) peaks. The dashed and solid lines represent the best fits of the corresponding experimental data to Eq. (1). The calculated dependence, in the absence of the DMI, is represented by the green dotted line.

The frequency nonreciprocity of the spin waves, which can be defined as $\Delta f = f(k) - f(-k)$, gives a measure of the strength of DMI present in a sample. As the second term on the RHS of Eq. (1) is an even function of $k$, $\Delta f$ can be expressed as

$$\Delta f = \frac{2\gamma Dk}{\pi M_S} \sin\phi. \qquad (2)$$

Based on the above equation, the frequency difference $\Delta f$, of corresponding counter-propagating SWs, was calculated as a function of in-plane angle $\phi$, using the fitted value of $D = 0.44$ mJ/m$^2$. The calculations were repeated for $\Delta f$ as a function of $\sin\phi$. As shown in Fig. 3, the calculated results, (sinusoidal and linear dependences, respectively) are in good accord with experiment. It is noteworthy that the fitted $D$ value, determined from an in-plane angular dependence of magnon frequencies, is the same as that obtained previously for the same sample by using a different experimental method, namely by measuring the magnon dispersion relations in the Damon-Eshbach geometry [16].

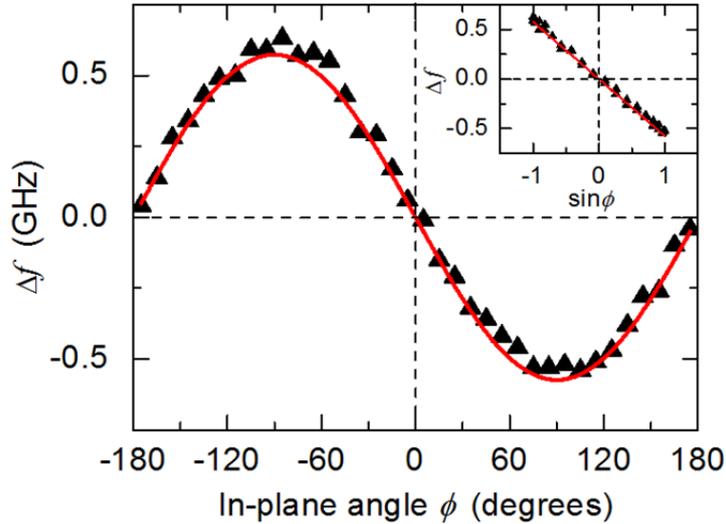

**FIG. 3.** Variations of the frequency difference $\Delta f$, of counter-propagating spin waves (corresponding to those shown in Fig. 2) of the Pt/Co/Ni film, with in-plane angle $\phi$, and $\sin\phi$ (inset). The triangles denote the measured data, while the solid line represents the calculated variation using Eq. (2), based on the fitted DMI constant $D = 0.44$ mJ/m$^2$.

It should be pointed out that, as one of our objectives is to verify the predicted in-plane angular dependence of the nonreciprocal SW propagation of Ref. 22, we have fitted the dependence of the SW frequencies $f$, to the fundamental Eq. (1), which involves the four parameters $M_S$, $\gamma$, $H_U$ and $J$. In practice, however if the aim is to just experimentally extract the DMI constant, one could use the derived Eq. (2) for $\Delta f$ as it involves fewer parameters, namely $M_S$ and $\gamma$. But the trade-off is that the experimental uncertainty for $\Delta f$ is larger than that for the individual frequencies $f$.

In conclusion, using Brillouin spectroscopy, we have experimentally verified the prediction, by Cortés-Ortuño and Landeros [22], of the in-plane angular dependence of the frequencies of spin waves propagating on a Pt/Co/Ni film, as well as the sinusoidal angular dependence of the frequency nonreciprocity $\Delta f$. Additionally, this study yielded a value of the DMI constant that is the same as that obtained by Di *et al*. from measurements of the spin wave dispersion relations [16]. Hence, our approach offers an alternative magnon-dynamics method, based on Brillouin spectroscopy, for ascertaining the interfacial DMI constants of multilayer thin films, an important parameter in the study of these interactions and novel phenomena induced by them, such as magnetic skyrmions.

This research was funded by the Ministry of Education, Singapore under Academic Research Fund Grant No. R144-000-340-112 and the National Research Foundation, Prime Minister's Office, Singapore under its Competitive Research Programme (CRP Award No. NRF-CRP12-2013-01).